\documentclass[aps,prl,twocolumn,superscriptaddress,floatfix,longbibliography]{revtex4-1}
\usepackage[dvipsnames]{xcolor}
\usepackage{amssymb,amsmath,amsfonts,bm,slashed,soul,ragged2e,graphicx,epstopdf,hyperref,array}
\usepackage[utf8]{inputenc}

\usepackage{colortbl,color,epsfig,subfigure,caption}
\enlargethispage{\baselineskip}
\definecolor{steelblue}{RGB}{25,25,112}
\definecolor{dullblue}{rgb}{0,0.298,0.49}
\definecolor{darkred}{rgb}{0.545,0,0}
\definecolor{blue2}{cmyk}{1, 0.1, 0.1, 0}
\hypersetup{colorlinks,linkcolor={darkred},citecolor={blue},urlcolor={dullblue}}
\DeclareCaptionJustification{justified}{\justifying}
\everymath{\displaystyle} 

\usepackage{comment}
\usepackage{braket}
\usepackage{amsmath}
\usepackage{soul}
\usepackage{makecell}
\usepackage{verbatim}

\newcommand{\beq}{\begin{equation}}
\newcommand{\eeq}{\end{equation}}
\newcommand{\bea}{\begin{eqnarray}}
\newcommand{\eea}{\end{eqnarray}}

\newcommand{\gsim}{\lower.7ex\hbox{$\;\stackrel{\textstyle>}{\sim}\;$}}
\newcommand{\lsim}{\lower.7ex\hbox{$\;\stackrel{\textstyle<}{\sim}\;$}}

\newcommand{\be}{\begin{equation}}
\newcommand{\ee}{\end{equation}}
\newcommand{\ba}{\begin{eqnarray}}
\newcommand{\ea}{\end{eqnarray}}

\newcommand{\D}{\mathrm{d}}

\newcommand{\mi}{\mathrm{i}}

\newcommand{\ghz}{\text{GHz}}

\newcommand{\vvir}{v_{\text{vir}}}

\RequirePackage[normalem]{ulem}

\begin{document}

\title{First Scan Search for Dark Photon Dark Matter with a Tunable Superconducting Radio-Frequency Cavity}

\author{Zhenxing Tang}
\email{The two authors have contributed equally.}
\affiliation{School of Physics and State Key Laboratory of Nuclear Physics and Technology, Peking University, Beijing 100871, China}
\affiliation{Beijing Laser Acceleration Innovation Center, Huairou, Beijing, 101400, China}

\author{Bo Wang}
\email{The two authors have contributed equally.}
\affiliation{
International Centre for Theoretical Physics Asia-Pacific,
University of Chinese Academy of Sciences, 100190 Beijing, China}

\author{Yifan Chen}
\affiliation{
Niels Bohr International Academy, Niels Bohr Institute, Blegdamsvej 17, 2100 Copenhagen, Denmark}

\author{Yanjie Zeng}
\affiliation{
CAS Key Laboratory of Theoretical Physics, Institute of Theoretical
Physics, Chinese Academy of Sciences, Beijing 100190, China}
\affiliation{School of Physical Sciences, University of Chinese Academy of Sciences, No. 19A Yuquan Road, Beijing 100049, China}
\author{Chunlong Li}
\affiliation{
CAS Key Laboratory of Theoretical Physics, Institute of Theoretical
Physics, Chinese Academy of Sciences, Beijing 100190, China}
\author{Yuting Yang}
\affiliation{
CAS Key Laboratory of Theoretical Physics, Institute of Theoretical
Physics, Chinese Academy of Sciences, Beijing 100190, China}
\affiliation{School of Physical Sciences, University of Chinese Academy of Sciences, No. 19A Yuquan Road, Beijing 100049, China}

\author{Liwen Feng}
\affiliation{School of Physics and State Key Laboratory of Nuclear Physics and Technology, Peking University, Beijing 100871, China}
\affiliation{Institute of Heavy Ion Physics, Peking
University, Beijing 100871, China}

\author{Peng Sha}
\affiliation{Institute of High Energy Physics, Chinese Academy of
Sciences, Beijing 100049, China}
\affiliation{
Key Laboratory of Particle Acceleration Physics and
Technology, Chinese Academy of Sciences, Beijing 100049,
China}
\affiliation{Center for Superconducting RF and Cryogenics, Institute of
High Energy Physics, Chinese Academy of Sciences,
Beijing 100049, China}
\author{Zhenghui Mi}
\affiliation{Institute of High Energy Physics, Chinese Academy of
Sciences, Beijing 100049, China}
\affiliation{
Key Laboratory of Particle Acceleration Physics and
Technology, Chinese Academy of Sciences, Beijing 100049,
China}
\affiliation{Center for Superconducting RF and Cryogenics, Institute of
High Energy Physics, Chinese Academy of Sciences,
Beijing 100049, China}
\author{Weimin Pan}
\affiliation{Institute of High Energy Physics, Chinese Academy of
Sciences, Beijing 100049, China}
\affiliation{
Key Laboratory of Particle Acceleration Physics and
Technology, Chinese Academy of Sciences, Beijing 100049,
China}
\affiliation{Center for Superconducting RF and Cryogenics, Institute of
High Energy Physics, Chinese Academy of Sciences,
Beijing 100049, China}
\author{Tianzong Zhang}
\affiliation{School of Physics and State Key Laboratory of Nuclear Physics and Technology, Peking University, Beijing 100871, China}
\author{Yirong Jin}
\affiliation{Beijing Academy of Quantum Information Sciences, Beijing 100193, China}

\author{Jiankui Hao}
\affiliation{School of Physics and State Key Laboratory of Nuclear Physics and Technology, Peking University, Beijing 100871, China}
\affiliation{Institute of Heavy Ion Physics, Peking
University, Beijing 100871, China}

\author{Lin Lin}
\affiliation{School of Physics and State Key Laboratory of Nuclear Physics and Technology, Peking University, Beijing 100871, China}
\affiliation{Institute of Heavy Ion Physics, Peking
University, Beijing 100871, China}
\author{Fang Wang}
\affiliation{School of Physics and State Key Laboratory of Nuclear Physics and Technology, Peking University, Beijing 100871, China}
\affiliation{Institute of Heavy Ion Physics, Peking
University, Beijing 100871, China}
\author{Huamu Xie}
\affiliation{School of Physics and State Key Laboratory of Nuclear Physics and Technology, Peking University, Beijing 100871, China}
\affiliation{Institute of Heavy Ion Physics, Peking
University, Beijing 100871, China}
\author{Senlin Huang}
\affiliation{School of Physics and State Key Laboratory of Nuclear Physics and Technology, Peking University, Beijing 100871, China}
\affiliation{Institute of Heavy Ion Physics, Peking
University, Beijing 100871, China}

\author{Jing Shu}
\email{Corresponding author: jshu@pku.edu.cn}
\affiliation{School of Physics and State Key Laboratory of Nuclear Physics and Technology, Peking University, Beijing 100871, China}
\affiliation{Beijing Laser Acceleration Innovation Center, Huairou, Beijing, 101400, China}
\affiliation{Center for High Energy Physics, Peking University, Beijing 100871, China}

\collaboration{SHANHE Collaboration}

\begin{abstract}
Dark photons have emerged as promising candidates for dark matter, and their search is a top priority in particle physics, astrophysics, and cosmology. We report the first use of a tunable niobium superconducting radio-frequency cavity for a scan search of dark photon dark matter with innovative data analysis techniques. We mechanically adjusted the resonant frequency of a cavity submerged in liquid helium at a temperature of $2$ K, and scanned the dark photon mass over a frequency range of $1.37$ MHz centered at $1.3$ GHz. Our study leveraged the superconducting radio-frequency cavity's remarkably high quality factors of approximately $10^{10}$, resulting in the most stringent constraints to date on a substantial portion of the exclusion parameter space on the kinetic mixing coefficient  $\epsilon$ between dark photons and electromagnetic photons, yielding a value of $\epsilon < 2.2 \times 10^{-16}$.

\end{abstract}
\date{\today}

\maketitle

\noindent{\it Introduction.}
---The quest for new physics in fundamental research has required increasingly precise measurements in recent years, specifically in detecting feeble signals from dark matter, whose existence is of utmost importance in understanding the structure and evolution of the Universe. Ultralight bosons, such as axions ~\cite{Preskill:1982cy,Abbott:1982af,Dine:1982ah} and dark photons~\cite{Nelson:2011sf,Arias:2012az}, which are predicted in many extra dimension or string-inspired models~\cite{Svrcek:2006yi,Abel:2008ai,Arvanitaki:2009fg,Goodsell:2009xc}, have become notable examples of such candidates. A dark photon, a hypothetical particle from beyond the standard model of particle physics, serves as the hidden gauge boson of a U(1) interaction. Through a small kinetic mixing, dark photons can interact with ordinary photons, thus providing one of the simplest extensions to the standard model.

The detection of ultralight dark photon dark matter (DPDM) capitalizes on the tiny kinematic mixing, which contributes to weak localized effective electric currents and enables experimental probing of these elusive particles. Various search techniques for DPDM have been employed, such as dish antennas~\cite{Horns:2012jf,FUNKExperiment:2020ofv,Ramanathan:2022egk}, geomagnetic fields~\cite{Fedderke:2021aqo,Fedderke:2021rrm}, atomic spectroscopy~\cite{Berger:2022tsn}, radio telescopes~\cite{An:2022hhb}, and atomic magnetometers~\cite{Jiang:2023jhl}. Additionally, due to similarities with axion detection~\cite{Sikivie:1983ip,Sikivie:1985yu,Sikivie:2013laa,Chaudhuri:2014dla,Kahn:2016aff}, axion-photon coupling constraints have been reinterpreted to set bounds on the kinetic mixing coefficient of dark photons~\cite{Ghosh:2021ard,Caputo:2021eaa}.

Haloscopes serve as a crucial tool for detecting ultralight dark matter. In these devices, the ultralight dark matter field is converted into electromagnetic signals within a cavity. The ongoing rapid advancements in quantum technology are anticipated to significantly bolster the sensitivity of these experimental setups~\cite{WLC, Chen:2021bgy,Wurtz:2021cnm,Jiang:2022vpm, Zheng:2016qjv,Malnou:2018dxn,HAYSTAC:2020kwv,Lehnert:2021gbj,HAYSTAC:2023cam,Dixit:2020ymh,Agrawal:2023umy}. Superconducting radio-frequency (SRF) cavities in accelerators~\cite{Padamsee:2017ohf} boast exceptionally high quality factors, reaching $Q_0 > 10^{10}$, allowing for the accumulation of larger electromagnetic signals and reduced noise levels~\cite{Dixit:2020ymh, Cervantes:2022gtv, Romanenko:2023irv,Agrawal:2023umy}. Unlike axion detection, DPDM detection does not require a magnetic field background, enabling the full potential of superconducting cavities to be exploited. Notably, the sensitivity to the kinetic mixing coefficient of the dark photon can experience enhancement by a factor of $Q_0^{-1/4}$ in scenarios where $Q_0 > Q_{\rm DM}$~\cite{Cervantes:2022gtv}. Here, $Q_{\rm DM} \approx 10^6$ characterizes the frequency spectrum of ultralight bosonic fields originating from a virialized velocity dispersion of $\sim 10^{-3}$\,c.

Exploring the extensive and as yet unexplored domain within the DPDM parameter space necessitates a detector capable of systematically scanning the mass window. This imperative calls for the incorporation of a frequency tuning structure, which marks an advancement over prior investigations focused on individual bins~\cite{Dixit:2020ymh,Cervantes:2022gtv,Agrawal:2023umy}. An SRF tuning structure was recently employed in a ``light-shining-through-wall" experiment for conducting broadband searches concerning dark photons~\cite{Romanenko:2023irv}. In this study, for the first time, we conducted scan searches for DPDM by mechanically tuning the SRF cavity. Furthermore, a novel data analysis strategy tailored for the $Q_0 > Q_{\text{DM}}$ regime was employed. This approach allowed us to access the deepest region of DPDM interaction across a majority of the scanned mass window, covering a total span of $1.37$ MHz centered around a resonant frequency of $1.3$ GHz. This effort represents the inaugural run of the Superconducting cavity as High-frequency gravitational wave, Axion, and other New Hidden particle Explorer (SHANHE) collaboration.

\noindent{\it A tunable SRF cavity for dark photon dark matter.}
---Dark photon field, denoted as $A^{\prime}_\mu$, can kinetically mix with the electromagnetic photon $A_\mu$ with a form $\epsilon F_{\mu\nu}^\prime F^{\mu\nu}/2$, where $\epsilon$ is the kinetic mixing coefficient, and $F_{\mu\nu}^\prime$, $F^{\mu\nu}$ are the corresponding field tensors. When a coherently oscillating DPDM field is present within a cavity, it generates an effective current denoted as $\vec{J}_{\text{eff}} = \epsilon\, m_{A^\prime}^2 \vec{A}^{\prime}$ that pumps cavity modes, where  $m_{A^\prime}$ is the dark photon mass. The DPDM field consists of an ensemble sum of nonrelativistic vector waves, with frequencies distributed in a narrow window approximately equal to $m_{A^\prime}/(2\pi Q_{\rm DM})$ centered around $m_{A^\prime}/(2\pi)$.

If the resonant frequency $f_0$ of a cavity mode falls within the frequency band around $m_{A^\prime}/(2\pi)$, excitation of the electromagnetic field in that mode occurs, resulting in a signal power proportional to $\epsilon^2 m_{A^\prime} V C \rho_{A^\prime}$, where $V$ is the cavity volume, $C$ is the form factor that characterizes the overlap between a cavity mode and the DPDM wave function along a specific axis (see Supplemental Material for detail), and $\rho_{A^\prime}\approx 0.45$\,GeV/cm$^3$ is the local dark matter energy density.
On the other hand, both internal dissipation of the cavity and amplifiers introduce noise, $P_n = P_{\rm th} + P_{\rm amp}$. $P_{\rm th}$ represents the power of thermal noise in the cavity and is proportional to $T f_0/Q_0$, where $T$ is the temperature of the cavity.
The signal and thermal noise are distributed within the same bandwidth $\approx (\beta + 1)f_0/Q_0$ in the limit that the cavity's quality factor $Q_0$ is much greater than $Q_{\rm DM}$. Here $\beta$ is the dimensionless cavity coupling factor representing the ratio between the power transferred to the readout port and the internal dissipation. The noise from the amplifier is characterized by its effective noise temperature $T_{\rm amp}$. The spectrum of the amplifier noise is flat within a frequency range $\Delta f_0$, which is the range over which the cavity’s resonant frequency can be kept stable. Consequently, the amplifier noise dominates over the thermal noise when $T_{\rm amp}\approx T$.

The signal-to-noise ratio (SNR) of each scan step's search can be estimated by using the Dicke radiometer equation: $\text{SNR} =  \sqrt{ t_{\text{int}} \Delta f_{0}} P_{\rm sig}/P_n$~\cite{Dicke:1946glx}, where $t_{\text{int}}$ denotes the integration time. This estimation enables us to determine the level of sensitivity toward $\epsilon$,
\begin{equation}
\begin{aligned}
    \epsilon   \approx 2.8 \times 10^{-16} &\left( \frac{10^{10}}{Q_0}\right)^{\frac{1}{4}} \left( \frac{\xi}{100} \right)^{\frac{1}{4}} \left( \frac{ 4\, \textrm{L}}{V} \right)^{\frac{1}{2}} \left( \frac{0.5}{C}\right)^{\frac{1}{2}} \\
    &  \left( \frac{100\, \text{s}}{t_{\rm int}}\right)^{\frac{1}{4}}  \left( \frac{1.3\, \ghz}{f_0}\right)^{\frac{1}{4}} \left( \frac{T_{\rm amp}}{3\, \text{K}}\right)^{\frac{1}{2}},
\end{aligned}
\label{SEp}
\end{equation}
where $\xi \equiv \Delta f_{0} Q_0/ f_0$, and we require SNR\,$=1.64$, and take $\beta \approx 1$, and $T \approx T_{\rm amp}$, as calibrated in this study, and $\rho_{A^\prime} = 0.45$ GeV/cm$^3$. Equation\,(\ref{SEp}) shows that high quality factors improve sensitivity to $\epsilon$, as $\epsilon \propto Q_0^{-1/4}$. SRF cavities are therefore powerful transducers for detecting DPDM~\cite{Cervantes:2022gtv}. 

\begin{figure}[t]
    \centering
    \includegraphics[width=0.45\textwidth]{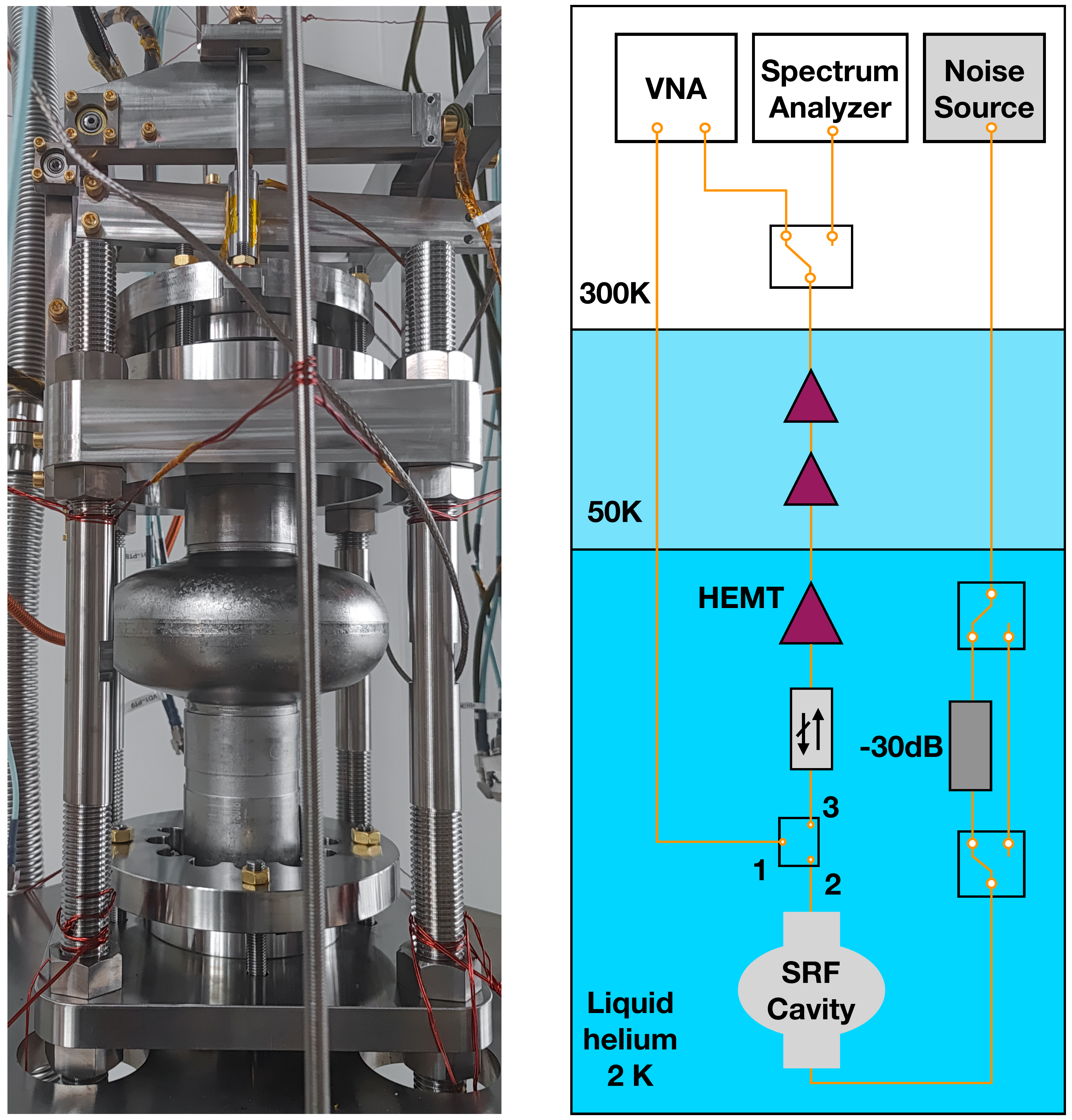}
    \caption{Left: single-cell SRF cavity equipped with frequency tuner. Right: schematic of the microwave electronics for DPDM searches. The VNA measures the net amplification factor $G_{\rm net}$ of the amplifier circuit consisting of an isolator, a HEMT amplifier, and two room-temperature amplifiers. The noise source and the spectrum analyzer calibrate the resonant frequencies $f_0^i$. The time-domain signals from the SRF, with sequential amplification, are finally recorded by the spectrum analyzer.
     }
    \label{SRFmodes}
\end{figure}

In this study, we used a single-cell elliptical niobium SRF cavity, as illustrated in Fig.\,\ref{SRFmodes}. The cavity has a volume $V \simeq 3.9$ L. We employ the ground mode TM$_{010}$ at $f_0 \simeq 1.3$ GHz, resulting in a form factor of $C \simeq 0.53$. To search DPDM within a reasonable mass range, it is imperative to scan the cavity at various resonant frequencies. To achieve this, a double lever frequency tuner~\cite{Pischalnikov:2015eye,Pischalnikov:2019iyu}, as depicted in Fig.\,\ref{SRFmodes}, was installed on the cavity. This tuner includes a stepper motor with a tuning resolution of approximately $10$ Hz, and a piezo actuator capable of fine-tuning at a level of $0.1$ Hz. A detailed schematic of this tuner is provided in the Supplemental Material. The cavity, along with the tuning apparatus and the experimental platform, has undergone extensive testing over several years~\cite{Mi:2014wwa,Liu:2016lzs,Zhou:2018vcm,Zhang:2022bmh,YANG20221354092,app13158618}.

\noindent{\it Experimental operation.}
Before carrying out DPDM searches, it is essential to calibrate the relevant cavity and amplifier parameters. 
All calibrated parameters and the corresponding uncertainties are presented in Table.\,\ref{ParaT}.
Both the volume of the cavity and the form factor of the TM$_{010}$ mode are  calculated numerically, with $< 1\%$ uncertainty for effective volume $V_{\rm eff}\equiv V\,C/3$. 
This uncertainty originates from the slight discrepancy between the simulated resonant frequency and the experimentally measured one, along with potential effects such as thinning due to acid pickling procedures.
Here, the factor of $1/3$ accounts for the random distribution of DPDM polarization.

We present the experimental setup in which the microwave electronics are depicted in the right panel of Fig.\,\ref{SRFmodes}. The cavity is positioned within a liquid helium environment at a temperature $T\simeq2$ K and is connected to axial pin couplers. The amplifier line consists of an isolator, which serves to prevent the injection of amplifier noise into the cavity, a high-electron mobility transistor (HEMT) amplifier, and two room-temperature amplifiers. Initially, we used a vector network analyzer (VNA) to measure the net amplification factor $G_{\rm net}$ of the amplifier circuit, which considers the sequential amplification and potential decays within the line. Next, we conducted decay measurements with a noise source that went through the cavity, the amplifier line, and the spectrum analyzer, to calibrate the cavity loaded quality factor, $Q_L\equiv Q_0/(\beta+1)$. The cavity coupling factor, $\beta$, was calibrated in combination with the results of the standard vertical test stand.

For each scan step, we used the noise source to calibrate the resonant frequency $f_0$ of the cavity by locating the peak of the power spectral density. This injected noise, featuring a spectrum wider than the cavity's bandwidth, serves as an effective stand-in for synthetic signals, ensuring that our data analysis procedures are well-suited for accurate signal detection. Immediately after calibration, we switched off the noise source and inserted a $30$ dB attenuator to prevent the external noise from entering the cavity. We then used the spectrum analyzer to record the time-domain signals from the SRF cavity and amplifiers. Each scan took $t_{\rm int} = 100$ s. After each scan, the value of $f_0$ was adjusted by approximately $1.3$ kHz and the calibration of $f_0$ was restarted. A total of $N_{\rm bin}=1150$ scans were conducted, covering a frequency range of approximately $1.37$\,MHz. The highest resonant frequency, denoted by $f_0^{\rm max}$, occurred when the frequency tuner was not applied. The calibration process for $G_{\rm net}$, $Q_L$, and $\beta$ was conducted multiple times during the whole scan process, with uncertainties given by the measurement deviation.

One key challenge of DPDM searches using SRF is to ensure any potential signal induced from DPDM is within the resonant bin, as $f_0$ may drift with time or oscillate due to microphonics effect~\cite{Pischalnikov:2019iyu,Romanenko:2023irv}.
To determine the stability range of $f_0$, denoted as $\Delta f_0$, we measured the drift of $f_0$ every $50$ scans, matching the integration time $t_{\rm int}$ of a single scan step, and also assessed the effect of microphonics over the same duration (see Supplemental Material). 
The microphonics effect produces a resonant frequency distribution with a root mean square of $\delta f_m^{\rm rms} = 4.1$ Hz, which is dominant over the drift with a maximum deviation of $1.5$ Hz. To account for any potential deviations in $f_0$, we conservatively selected $\Delta f_0$ to be $2.8\,\delta f_m^{\rm rms} \simeq 11.5$ Hz, taking into consideration an efficiency of $84\%$ for the recorded signal to optimize the SNR.

\begin{table}[t]
\begin{tabular}{l c c}
\hline
\hline
        & Value                 & Fractional Uncertainty  \\
\hline
$V_{\rm eff}\equiv V\,C/3$              & $693$ mL                 & $< 1\%$   \\
$\beta$          & $0.634\pm 0.014$                  & $1.4\%$  \\
$G_{\rm net}$              & $(57.30 \pm 0.14)$\,dB               & $3.1\%$ \\
$Q_L$            & $(9.092\pm 0.081)\times 10^9$  & $\slash$  \\
$f_{0}^{\rm max}$            & $1.299\,164\,379\,5$\,GHz               & $\slash$  \\
$\Delta f_0$     & $11.5$ Hz                 & $\slash$  \\
$t_{\text{int}}$ & $100$ s                  & $\slash$  \\
\hline
\hline
\end{tabular}
\caption{Calibrated parameters for SRF cavities and amplifiers used are shown, including their mean values, uncertainties and fractional uncertainties on DPDM-induced power, $F_j$.}
\label{ParaT}
\end{table}

\begin{figure}[htb]
    \centering
    \includegraphics[width=0.45\textwidth]{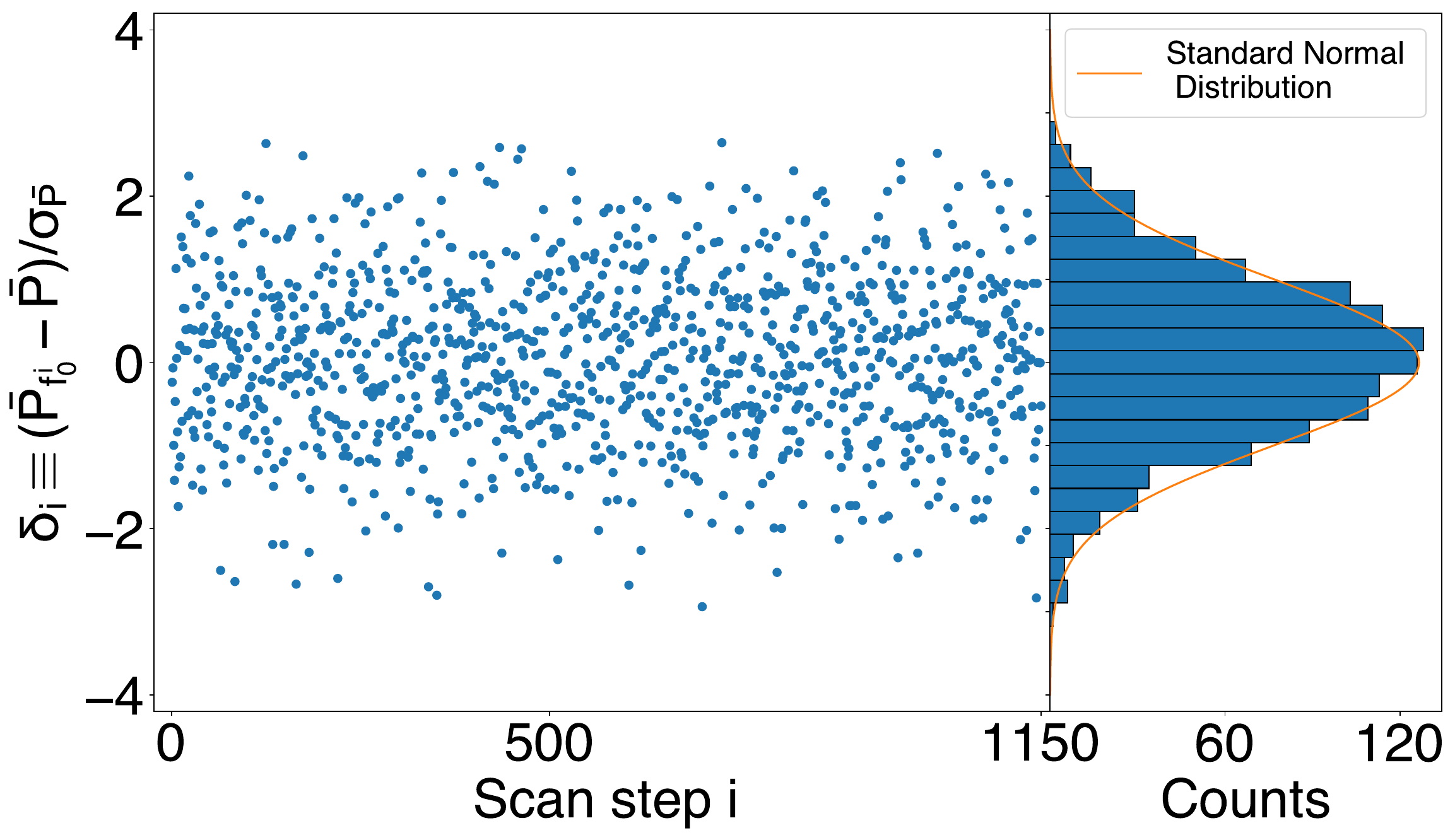}
    \caption{
    The blue dots show the normalized power excess $\delta_i \equiv (\bar{P}_{f_0^i}-\bar{P})/\sigma_{\bar{P}}$ at each scan step $i$. Its distribution is shown on the right panel, which can be well fit by a standard normal distribution.}
    \label{fig:2K}
\end{figure}

\noindent{\it Data analysis and constraints.}
---In this study, each scan was focused on the frequency bin centered at the resonant frequency $f_0$, which had a bandwidth of $\Delta f_0$. For every  scan, we obtained $N = t_{\rm int}\Delta f_0$ samples at the resonant bin and computed their average value and standard deviation. We checked the Gaussian noise property by ensuring that the ratio between these two values was close to $1$ at each step. The average values of different scans provided an indication of the total noise in each resonant bin. The amplifier noise, $P_{\rm amp}$, was found to be nearly constant over the entire frequency range tested. Furthermore, the subdominant thermal noise was observed to be linearly proportional to the resonant frequency, with a variation much smaller than the standard deviation. Therefore, we expected the noise in the resonant bins to be independent of the resonant frequency. To reduce the potential effects of environmental variation, such as helium pressure fluctuations and mechanical vibrations, we aggregated every $50$ contiguous bins to ensure environmental stability within each group. 
For each group, we computed a constant fit for different bins and presented the normalized power excess in Fig.~\ref{fig:2K}. 
The right panel of the figure shows a comparison between the counts of normalized power excess and the standard normal distribution to confirm its Gaussianity. No deviation over $3\sigma$ appears in any bin. Note that the scan steps do not progress in a strictly monotonic order by frequency, as continuously tuning the frequency in a single direction can induce additional drift of $f_0$. Monotonic progression is maintained only within groups of $50$ consecutive bins.

Compared to the analysis strategies employed by traditional haloscopes with $Q_0\ll Q_{\rm DM}$, our resonant bins cover only a fraction of the entire frequency band, $\Delta f_0 Q_{\rm DM}/f_0$. However, we can still test the DPDM with masses within this range and thereby maximize the scan rate. Furthermore, our simple fit function results in attenuation factor of $98\%$. This value is less suppressed when compared to low $Q_0$ experiments, where higher-order fitting functions are utilized to account for the frequency-dependent cavity response during each scan.

\begin{figure}[htb]
    \centering
    \includegraphics[width=0.48\textwidth]{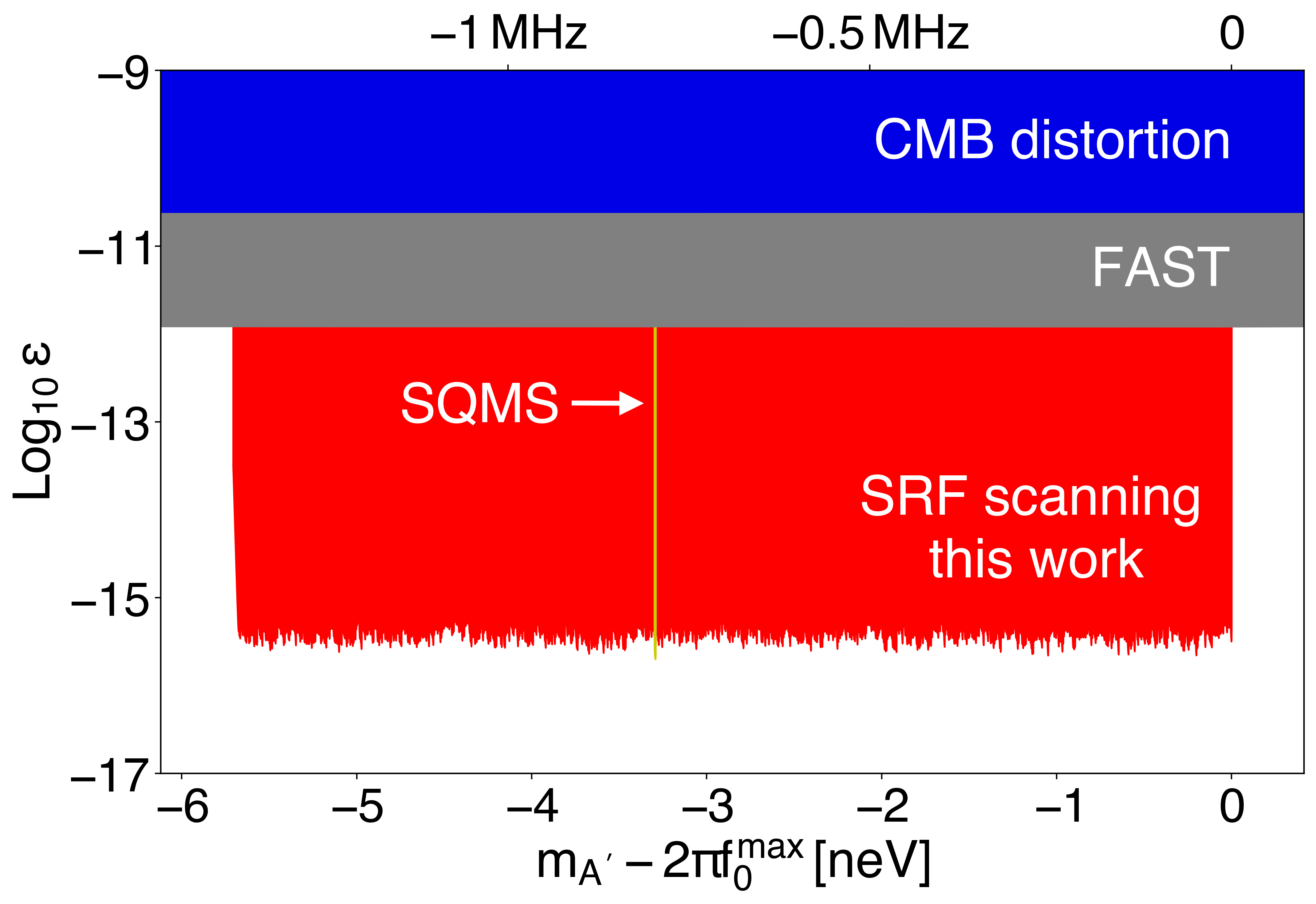}\\
       \includegraphics[width=0.49\textwidth]
       {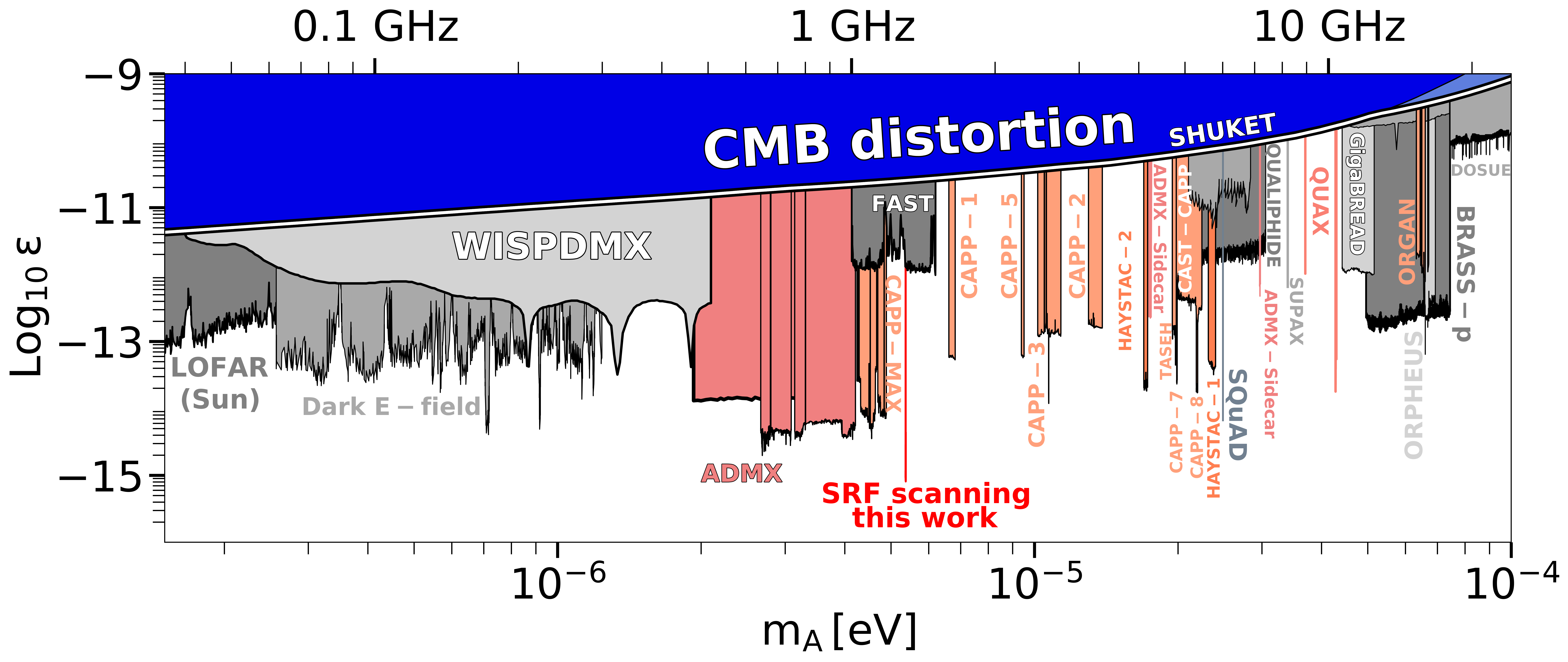} 
    \caption{Top: the $90\%$ exclusion on the kinetic mixing coefficient $\epsilon$ of DPDM based on SRF scan searches performed in this study (red). Other constraints including FAST radio telescope (gray)~\cite{An:2022hhb}, distortion of cosmic microwave background (blue)~\cite{Arias:2012az}, and SQMS prototype (yellow)~\cite{Cervantes:2022gtv}  are shown for comparison. Bottom: a comparison of our results within the broader context of existing constraints, adapted from \cite{AxionLimits}.}
    \label{fig:DPexclusion}
\end{figure}

There are two sources of uncertainty that affect the sensitivity toward DPDM searches. In addition to the fit uncertainty caused by Gaussian noise, there are also uncertainties in calibrated parameters that may contribute to a biased estimate for DPDM-induced signals. We present the measurement uncertainties of parameters ${ V_{\rm eff}, \beta, G_{\rm net} }$ and their corresponding fractional influences on signal power in Table\,\ref{ParaT} (see Supplemental Material for details). To compute the probability function for a potential DPDM signal, we multiply the contributions from different bins. However, because the DPDM width $\approx m_{A^\prime}/(2\pi Q_{\rm DM})$ is much larger than the narrow bandwidth $\Delta f_0$, we only consider the two nearby bins in practice. Figure\,\ref{fig:DPexclusion} shows the $90\%$ upper limits on the kinetic mixing coefficient $\epsilon$ for a given DPDM mass $m_{A^\prime}$. The high quality factor of SRF significantly boosts sensitivity, leading to the most stringent constraints compared to other limitations across a wide range of investigated masses. The reached sensitivity is well-estimated by Eq.\,(\ref{SEp}). For comparative analysis, we present the outcomes of a single-bin search conducted in SQMS~\cite{Cervantes:2022gtv} in the top panel. Both investigations utilized a conventional $1.3$ GHz elliptical cavity, yielding akin parameters encompassing $V_{\text{eff}}$, $f_0$, $\beta$, and $Q_L$. The primary distinction between our parameters and theirs lies in the bin size and integration time. Specifically, our $t_{\rm int}$ is 10 times shorter than theirs. We conservatively selected $\Delta f_0 = 11.5$ Hz, whereas their choice is only $0.15$ Hz. The bottom panel presents a comparison across a wider frequency range with other experiments, clearly demonstrating that SRF experiments achieve the deepest sensitivity.

\noindent{\it Conclusion.}
---In this study, we utilized a tunable single-cell $1.3$ GHz elliptical cavity to search for DPDM. Our findings establish the most stringent exclusion limit across a majority of the scanned mass window, achieving a depth of sensitivity of up to $\epsilon \sim 2.2\times 10^{-16}$. This result demonstrates that employing cavities with high quality factors significantly enhances the sensitivity toward the kinetic mixing coefficient of DPDM.
Our experiment presents the first scan results using a tunable SRF cavity, which covers a frequency range of $1.37$ MHz within DPDM's mass window, beginning from an initial resonant frequency of approximately $f_0^{\rm max} \simeq 1.299$ GHz.
Our scan steps are set at intervals corresponding to $10^{-6}$ of the resonant frequency, aligning with the dark matter bandwidth to optimize the scan rate. To investigate any potential excess from a suspicious signal, we can simply adjust the resonant frequency slightly away from the bin indicating excess. Conducting a comprehensive scan of the surrounding region allows for the reconstruction of the frequency spectrum of DPDM, providing valuable insights into the mechanisms of dark matter formation.

In the upcoming phase of our DPDM search, our foremost goals are to broaden the tuning range and boost sensitivity. We are in the process of designing a new tuning mechanism—a plunger tuner—that will adjust the beam pipe's end face at one end of the cavity. This adjustment is projected to enhance the tuning range to approximately $1/10$ of the resonant frequency. To further augment sensitivity, our strategy includes mitigating microphonics effects and diminishing amplifier noise, utilizing dilution refrigeration and nearly quantum-limited Josephson parametric amplifiers. Additionally, by employing coupled-cavity designs, we anticipate increasing the cavity volume tenfold while maintaining the same resonant frequency. With these advancements combined, we are optimistic about setting new constraints on the kinetic mixing coefficient $\epsilon$, potentially below $10^{-17}$.

The exceptionally high quality factors of SRF cavities open avenues for additional enhancements in detection sensitivity.
For example, coupling a single cavity mode to a multimode resonant systems with nondegenerate parametric interactions~\cite{WLC, Chen:2021bgy, Wurtz:2021cnm,Jiang:2022vpm} can broaden the effective bandwidth of each scan without losing sensitivity within it. One can also exploit squeezing technology~\cite{Zheng:2016qjv, Malnou:2018dxn, HAYSTAC:2020kwv, Lehnert:2021gbj,HAYSTAC:2023cam} or nondemolition photon counting~\cite{Dixit:2020ymh,Agrawal:2023umy} to go beyond the standard quantum limit. 
A network of DPDM detectors simultaneously measuring at the same frequency band will not only increase the sensitivity~\cite{Chen:2021bgy, Brady:2022bus}, but also reveal macroscopic properties and the microscopic nature of the DPDM sources, such as the angular distribution and polarization~\cite{Foster:2020fln, Chen:2021bdr}.

\hspace{5mm}
\begin{acknowledgements}

\noindent{\it \bfseries Acknowledgements.} 
We are grateful to Raphael Cervantes for useful discussions. We acknowledge the utilization of the Platform of Advanced Photon Source Technology R\&D. This work is supported by the National Key Research and Development Program of China under Grant No. 2020YFC2201501, and by the Munich Institute for Astro-, Particle, and BioPhysics (MIAPbP), which is funded by the Deutsche Forschungsgemeinschaft (DFG, German Research Foundation) under Germany´s Excellence Strategy – EXC-2094 – 390783311.
Y.C. is supported by VILLUM FONDEN (grant no. 37766), by the Danish Research Foundation, and under the European Union’s H2020 ERC Advanced Grant “Black holes: gravitational engines of discovery” grant agreement no. Gravitas–101052587, and by FCT (Fundação para a Ciência e Tecnologia I.P, Portugal) under project No. 2022.01324.PTDC. 
P.S. is supported by the National Natural Science Foundation of China under Grant No. 12075270.
J.S. is supported by Peking University under startup Grant No. 7101302974 and the National Natural Science Foundation of China under Grants No. 12025507, No. 12150015, and is supported by the Key Research Program of Frontier Science of the Chinese Academy of Sciences (CAS) under Grant No. ZDBS-LY-7003 and CAS project for Young Scientists in Basic Research YSBR-006.

\end{acknowledgements}

%

\appendix
\pagebreak
\widetext
\begin{center}
\textbf{\large Supplemental Materials: First Scan Search for Dark Photon Dark Matter with Superconducting Radio-frequency Cavity}
\end{center}
\setcounter{equation}{0}
\setcounter{figure}{0}
\setcounter{table}{0}
\makeatletter
\renewcommand{\theequation}{S\arabic{equation}}
\renewcommand{\thefigure}{S\arabic{figure}}
\renewcommand{\bibnumfmt}[1]{[#1]}
\renewcommand{\citenumfont}[1]{#1}

\section{Experimental Operation and Calibration}

Fig.\,1 of the maintext depicts the experimental setup employed in this study. The single-cell elliptical cavity was fitted with a double lever frequency tuner \cite{Pischalnikov:2015eye,Pischalnikov:2019iyu} and submerged in liquid helium at a temperature of $2$\,K. The cavity was coupled to an amplifier circuit, comprising an isolator, a high-electron mobility transistor (HEMT) amplifier (LNF-LNC0.6\_2A), and two room-temperature amplifiers (ZX60-P103LN+), via an axial pin coupler. The amplifier circuit was further linked to a spectrum analyzer (Rohde \& Schwarz FSV3030) or a vector network analyzer (VNA, Siglent SNA5054X). On the other end of the cavity, a highly undercoupled coupler was employed to enable noise injection.

The experimental operation included parameter calibration and data recording. The cavity volume and the form factor of the TM$_{010}$ mode were determined using the COMSOL numerical modeling software. Throughout the scan process, we performed multiple calibrations of $G_{\rm net}$, $Q_L$, and $\beta$ with the VNA and the standard vertical test stand (VTS). We calibrated the resonant frequency every scan step using a noise source. We also conducted a measurement for microphonics effect and resonant frequency drift tests every $50$ scan steps. The subsections below explain each calibration process in detail.

During each scan step, we configured the spectrum analyzer in the I/Q mode to record complex voltage signals from the SRF cavity and amplifiers in the time domain. We used the noise source to initially calibrate the resonant frequency $f_0$ of the cavity by identifying the peak of the power spectral density (PSD). After switching off the noise source, we applied a $30$\,dB attenuator to prevent external noise from entering the cavity and then recorded time-domain signals within an integration time of $t_{\rm int} = 100$ seconds. Before the end of each scan step, we conducted frequency calibration again to ensure that there was no frequency shift greater than $\Delta f_0$.

\begin{figure}[thb]
    \centering
    \includegraphics[width=0.33\columnwidth] {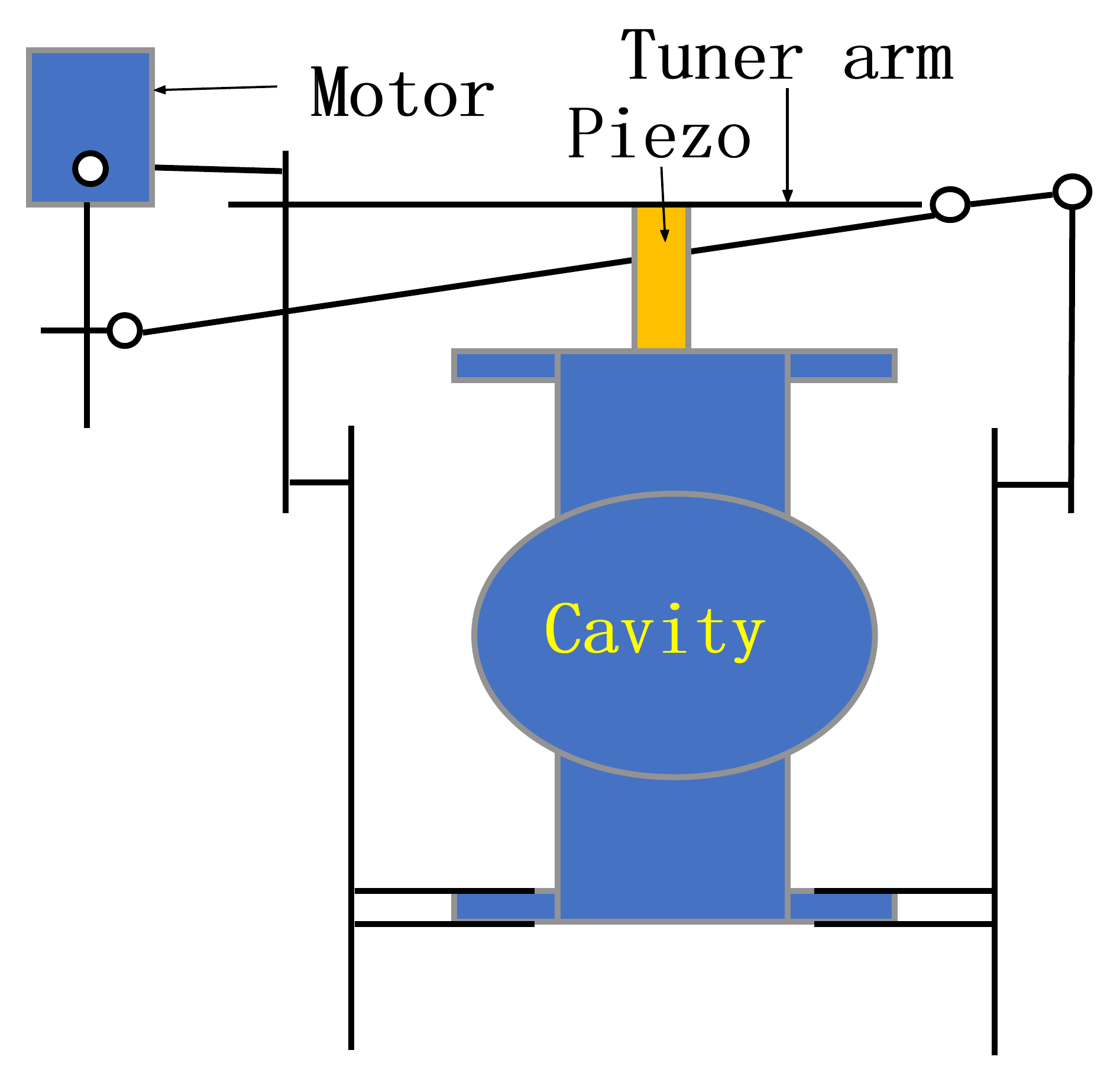}
    \includegraphics[width=0.55\columnwidth]{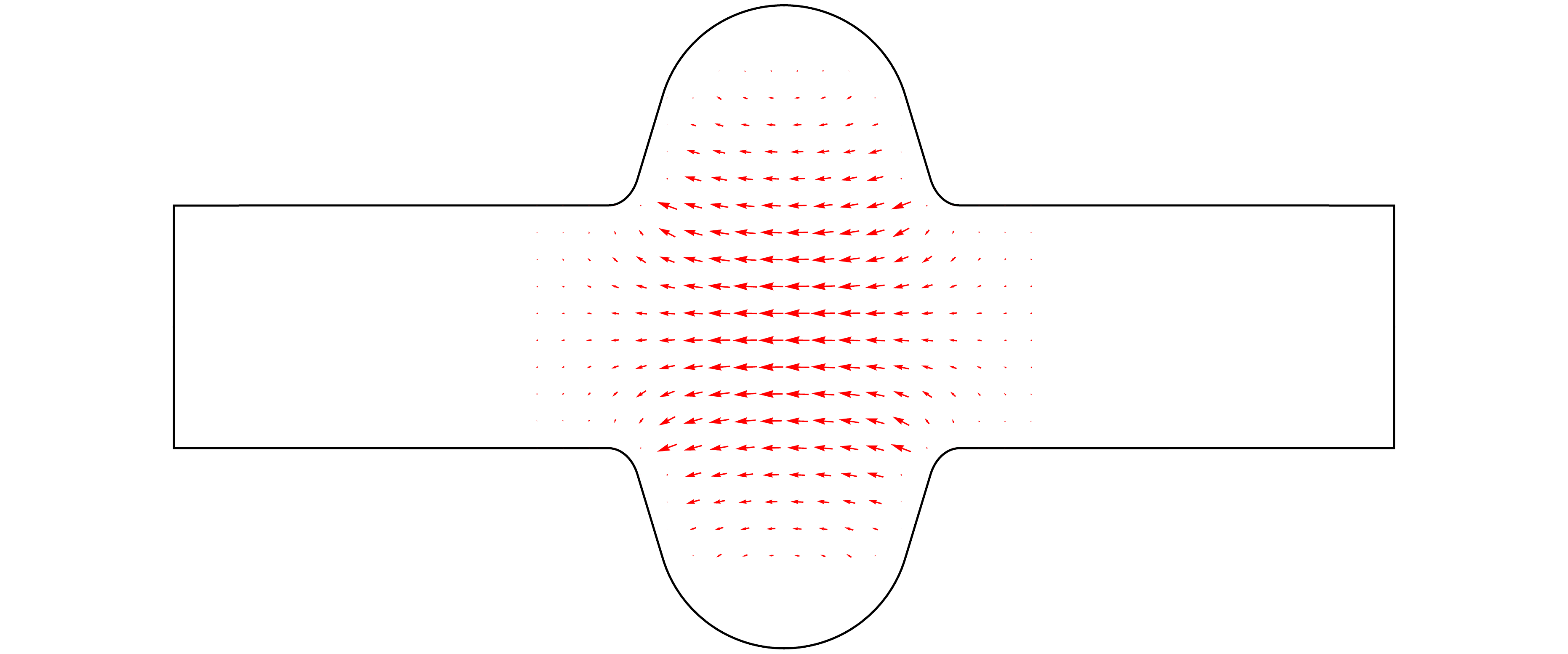}
    \caption{ Single-cell SRF cavity with ground mode TM$_{010}$ at $1.3$ GHz. 
    The electric field lines are shown in red, whose length is proportional to the electric field strength.
     }
    \label{SCC}
\end{figure}

After each scan, we adjusted the resonant frequency by approximately $1.3$\,kHz. {The left panel of Fig.\,\ref{SCC} illustrates the schematic of the tuning structure, which encompasses a motor, a piezo, a tuning arm, and a fixed arm positioned beneath the tuning arm. The motor, connected to the fixed arm, applies a downward force to the tuning arm. This force is transmitted to the piezo, and consequently conveyed to the flange located on the cavity, inducing compression of the cavity. The photograph in Fig.\,1 in the main text includes all the major components mentioned, except for the motor.} We used the spectrum mode of the spectrum analyzer to track the resonant frequency during tuning and subsequently reverted to the I/Q mode to initiate the next scan search.

The single-cell elliptical SRF cavity used in our study was created by combining two half-cell endcups that were originally used for 9-cell cavities \cite{Aune:2000gb}. To determine the cavity's volume, we used a corresponding analytic model and imported it into COMSOL, as illustrated in the right panle of Fig.\,\ref{SCC}. The cavity volume was calculated to be approximately $3.9$\,L using numerical spatial integration within the cavity.

The simulation provided us with the electric field distribution $\vec{E}_0(\vec{x})$ of the TM$_{010}$ mode. Fig.\,\ref{SCC} shows the electric field distribution in the plane that passes through the central axis of the cavity. The length of the red lines in the figure is proportional to the electric field strength. We calculated the form factor using the expression
\be
C \equiv \frac{1}{V} \left\vert {\int}_{V} \, \vec{E}_0 \, \D V\right\vert^2,\label{eqC}\ee
where the electric field is normalized to satisfy the condition $\int_V |\vec{E}_0|^2 \D V = 1$. A detailed discussion of Eq.\,(\ref{eqC}) is provided in the following section.

The DPDM-induced signal $P_{\rm sig}$ is proportional to the product of $V$ and $C$, both of which are determined from numerical simulations. We introduce the effective volume $V_{\rm eff}\equiv V\,C/3$, where the factor of $1/3$ accounts for the random distribution of DPDM polarization. The uncertainty of $V_{\rm eff}$ arises from several sources. Firstly, the cavity has undergone several acid pickling procedures in the past, which may have caused a potential thinning of the inner cavity wall by $\mathcal{O}(100)\,\mu$m, in comparison with its original design. This contributes to an uncertainty in $V_{\rm eff}$ of less than $1\%$. Secondly, tuning the resonant frequency in a range of $\mathcal{O}(1)$\,MHz may lead to an uncertainty of less than  $0.1\%$. Finally, we ensured the numerical simulation converged. Hence, we conservatively set the uncertainty of $V_{\rm eff}$ to be $1\%$.

\subsection{Calibration of $G_{\rm net}$} 
To determine the net amplification factor $G_{\rm net}$ of the amplifier circuit, we performed the following calibration procedure.  Initially, ports $1$ and $3$ of the cryogenic matrix switch in Fig.\,1 of the maintext were connected, and then port $1$ was linked to the VNA to establish a feedback loop between the amplifier circuit and the VNA. We measured the amplification factor of the loop, which we refer to as $G_{\rm loop}$. It should be noted that $G_{\rm loop}$ differed from $G_{\rm net}$ due to the attenuation resulting from the cable connecting the cryogenic matrix switch and the VNA. We refer to this attenuation as $G_{\rm cable}<0$. To determine $G_{\rm cable}$ using the VNA, we connected ports $1$ and $2$ in the off-resonant region of the cavity, which corresponds to the total reflection mode. Thus, the value of $G_{\rm net}$ is obtained by adding the magnitudes of $G_{\rm loop}$ and $|G_{\rm cable}|$ together. It is possible that the cable connecting the cryogenic matrix switch and the cavity could introduce additional insertion loss, which was not accounted for in the measurements. However, we can confidently neglect the impact of the cable due to its superconducting nature.

\subsection{Calibration of $Q_L$ and $\beta$.} The rate at which cavity mode decays is proportional to the inverse of its quality factor. The loaded quality factor, $Q_L$, which quantifies the inverse of the total energy loss rate, is determined by contributions from both intrinsic loss and energy extraction, as given by
\begin{equation}
    \frac{1}{Q_L}=\frac{1}{Q_0}+\frac{1}{Q_{\rm ext}},
\end{equation}
where $Q_0$ and $Q_{\rm ext}$ represent the intrinsic and external quality factors, respectively. As $Q_L$ is dependent on the field strength within the cavity \cite{Romanenko:2017nkv}, it was calibrated using a noise source that excited low field strength, thereby enabling it to converge to a value where only thermal noise exists in the cavity. Specifically, after turning off the noise source, the excited field decayed exponentially, i.e.,
\be
    P(t)=P(t_0)e^{-(t-t_0)/\tau}
\ee
 where $\tau$ represents the decay time. To determine the quality factor ($Q_L$) of the decay, the value of $\tau$ was fitted to yield $Q_L = 2\pi f_0\tau$.

The cavity coupling factor $\beta$ is defined as the ratio between the energy delivered to the readout antenna and the internal dissipation, i.e.,
\be \beta = \frac{Q_L}{Q_{\rm ext}-Q_L}.\ee
 Therefore, the measurement of $Q_{\rm ext}$, whose value is independent of the field strength inside the cavity, can be used to determine $\beta$. This measurement can be carried out using the standard vertical test stand \cite{Melnychuk:2014pka} involving both the measurement of forward and reflected power and decay measurement.

\subsection{Resonant Frequency Calibration and Stability.} 
\begin{figure}[htb]
    \centering
    \includegraphics[width=0.4\columnwidth] {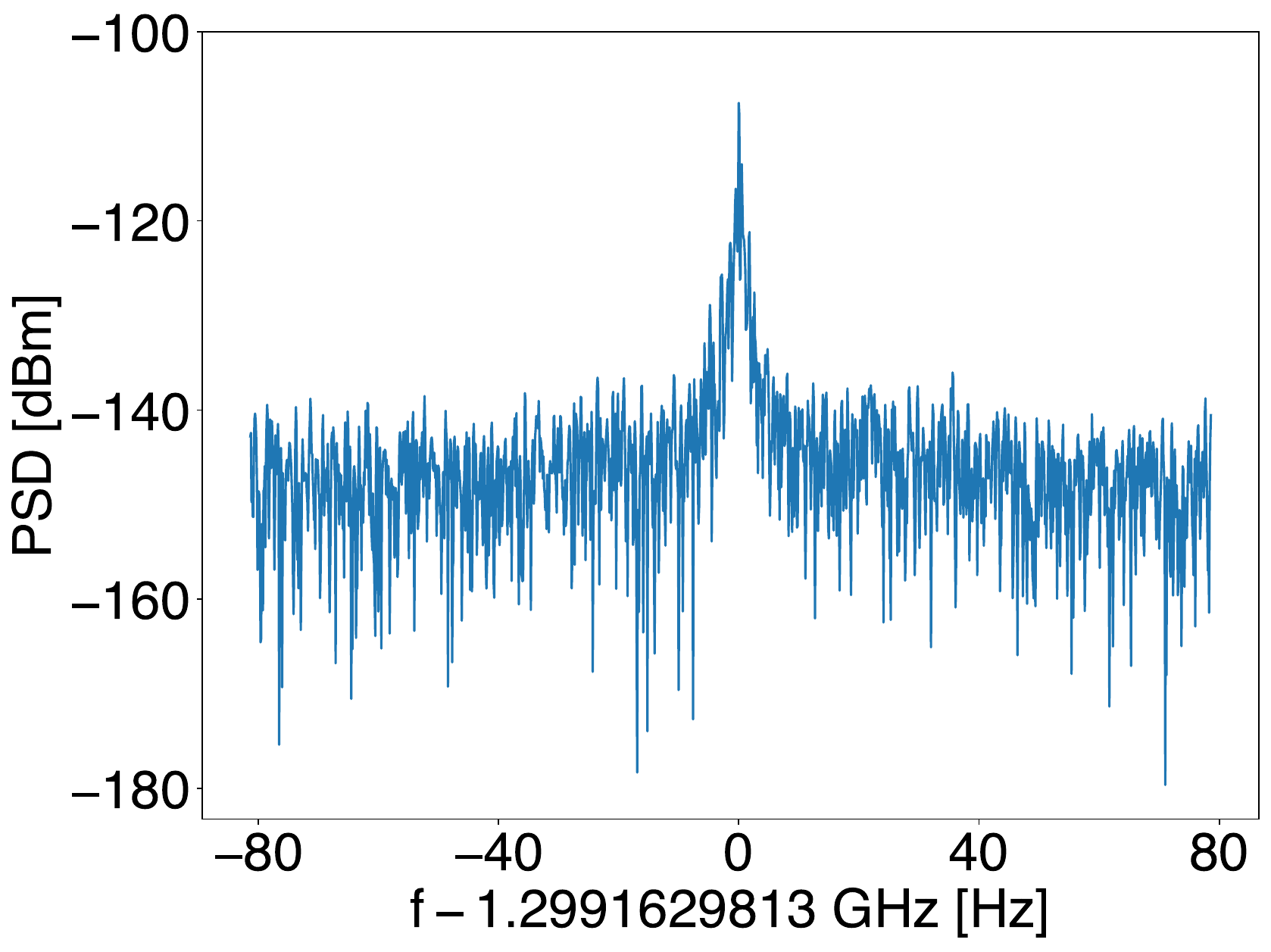}  
    \caption{An example of PSD injected by a noise source to calibrate the resonant frequency by finding the peak with an integration time of $10$ seconds.
     }
    \label{Fshift}
\end{figure}

Calibration of the resonant frequency is essential in the search for DPDM using SRF cavities, due to their high quality factor and narrow bandwidth. We accomplished this by injecting a broadband noise source into the cavity and recording the signal with a spectrum analyzer for $10$ seconds. We then selected the frequency bin with the peak power spectral density (PSD) as the resonant frequency. An example of the PSD is presented in Fig.\,\ref{Fshift}. Following each calibration, we switched off the noise source and waited for the excited field to decay until the cavity was dominated by noise. Each data recording began immediately thereafter.

\begin{figure}[htb]
    \centering
    \includegraphics[width=0.44\columnwidth]{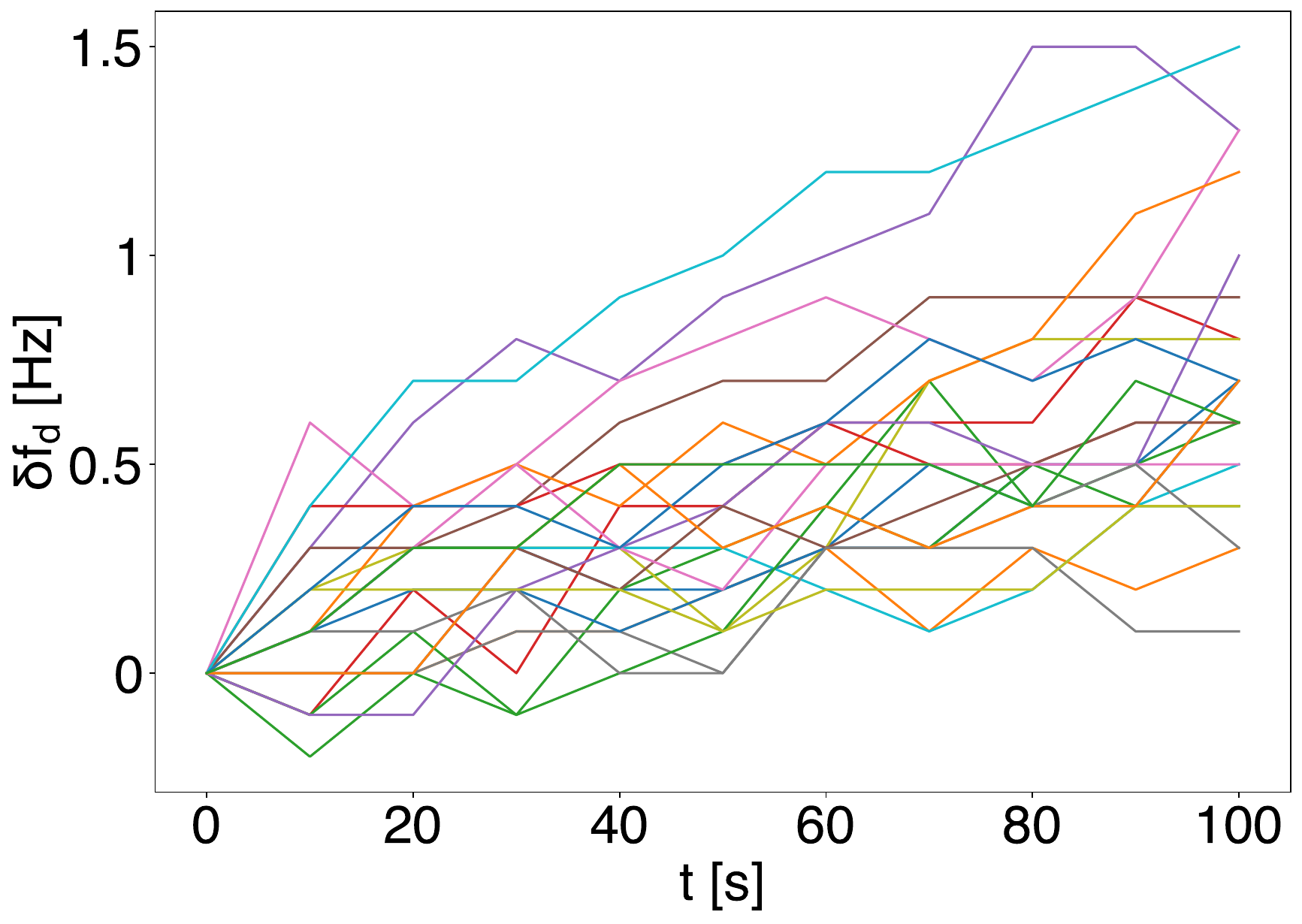} \qquad
    \includegraphics[width=0.425\columnwidth]{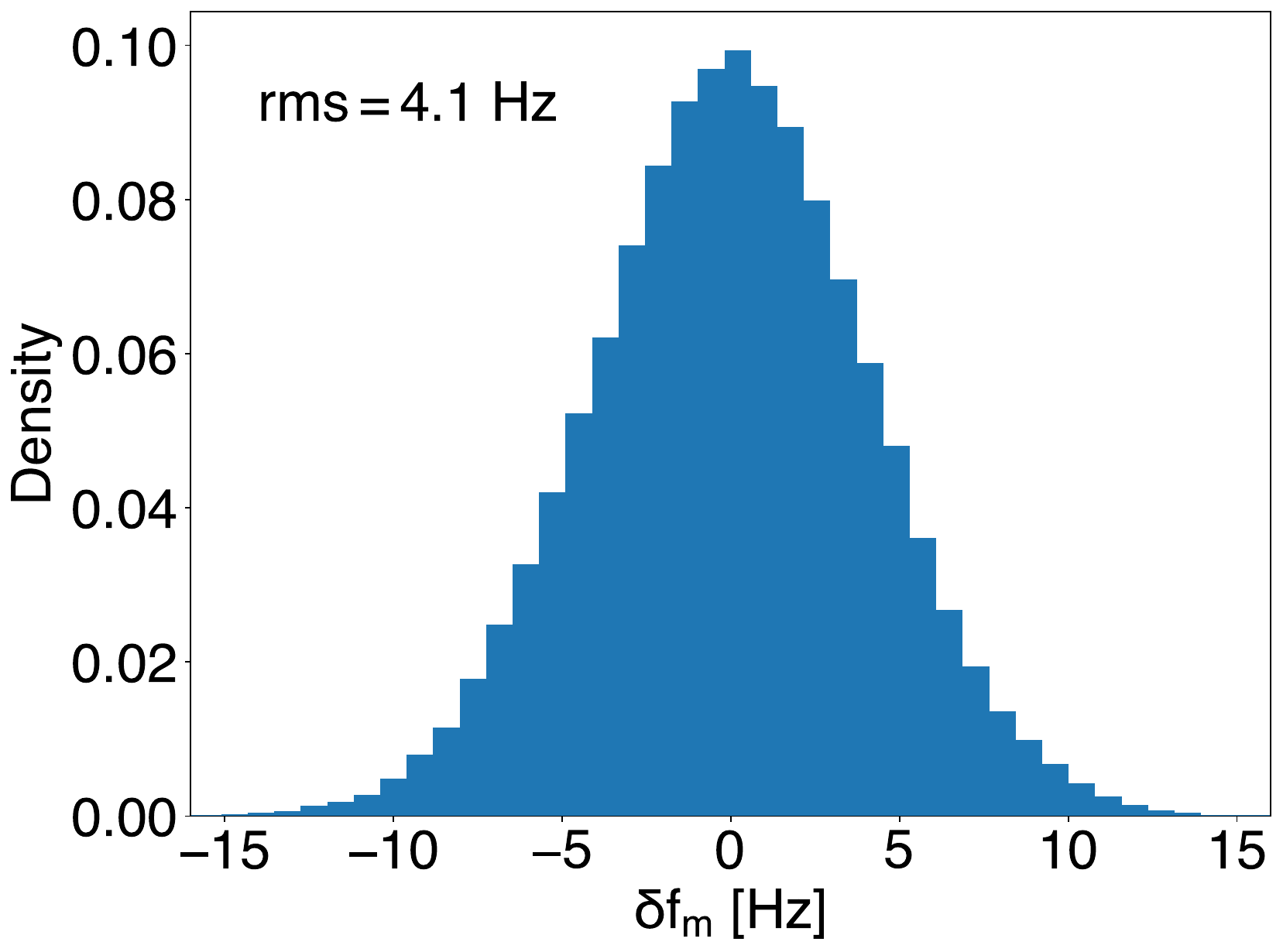}
    \caption{ Left: 
  measurements of resonant frequency drift for a duration of $100$ seconds. Every colored line represents a test that was conducted every $50$ scan steps.
  Right: histogram of resonant frequency's oscillatory deviation.}
    \label{mph}
\end{figure}

The stability analysis included both frequency drift and microphonics effect. 
For every group of $50$ scan steps, we tested the drift of the resonant frequency for $100$ seconds, which is equivalent to the data recording time for one scan.   The drift of $f_0$, denoted by $\delta f_d$, is presented in the left panel of Fig.\,\ref{mph}. As we selected the peak of the PSD as the resonant frequency bin every $10$ seconds, the $100$-second interval included the $10$-step evolution of $\delta f_d$. In most cases, we observed a gradual increase over time due to the resistance of the cavity to mechanical deformation. The maximal value of the frequency drift, $\delta f_d^{\rm max}$, is $1.5$\,Hz.

On the other hand, the microphonics effect that leads to oscillatory deviation of the resonant frequency \cite{Pischalnikov:2019iyu} does not necessarily reflect on the peak position of PSDs with $10$-second intervals. 
Instead, we employed the Digital Phase-Locked Loops (PLL) system available in the VTS to evaluate the microphonics \cite{Pischalnikov:2019iyu}. This system introduces a coherent field into the SRF cavity, and the oscillatory deviation of the resonant frequency is reflected in the change rate of the relative phase, given by $2\pi\,\delta f_m =\D \phi/\D t$. In the right panel of Fig.\,\ref{mph}, we present the results of the microphonics test for a $100$-second interval.
The histogram of $\delta f_m$ indicates the dominance of the drift effect, which follows a Gaussian distribution with a root mean square (rms) value of $\delta f_m^{\rm rms}=4.1$\,Hz. We conservatively assume that the accumulation of DPDM signals follows the same Gaussian distribution, with an efficiency $\eta_{\rm bin}$ given by $\eta_{\rm bin}=\text{erf}\left(\Delta f_0/(2\sqrt{2}\delta f_m^{\rm rms})\right)$, where $\text{erf}(x)$ represents the error function. By maximizing the signal-to-noise ratio (SNR) $\propto \eta_{\rm bin}/\sqrt{\Delta f_0}$, the choice of $\Delta f_0$ can be optimized, which results in $\Delta f_0 \simeq 2.8\,\delta f_m^{\rm rms}\simeq 11.5$\,Hz and $\eta_{\rm bin} \simeq 84\%$.

\section{Data Characterization and Analysis}
\begin{figure}[thb]
    \centering
 \includegraphics[width=0.42\columnwidth]{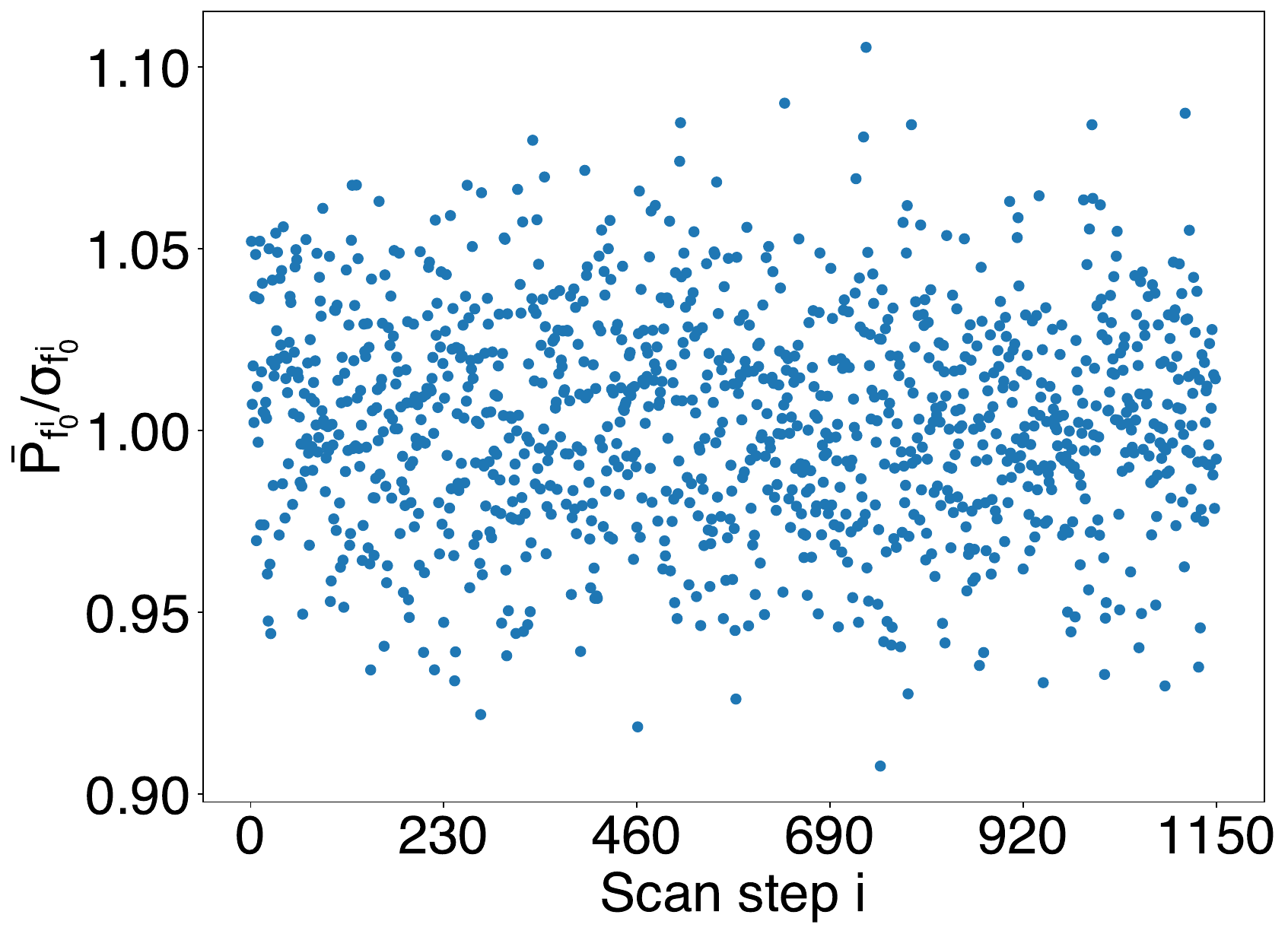} \qquad  \includegraphics[width=0.494\columnwidth]{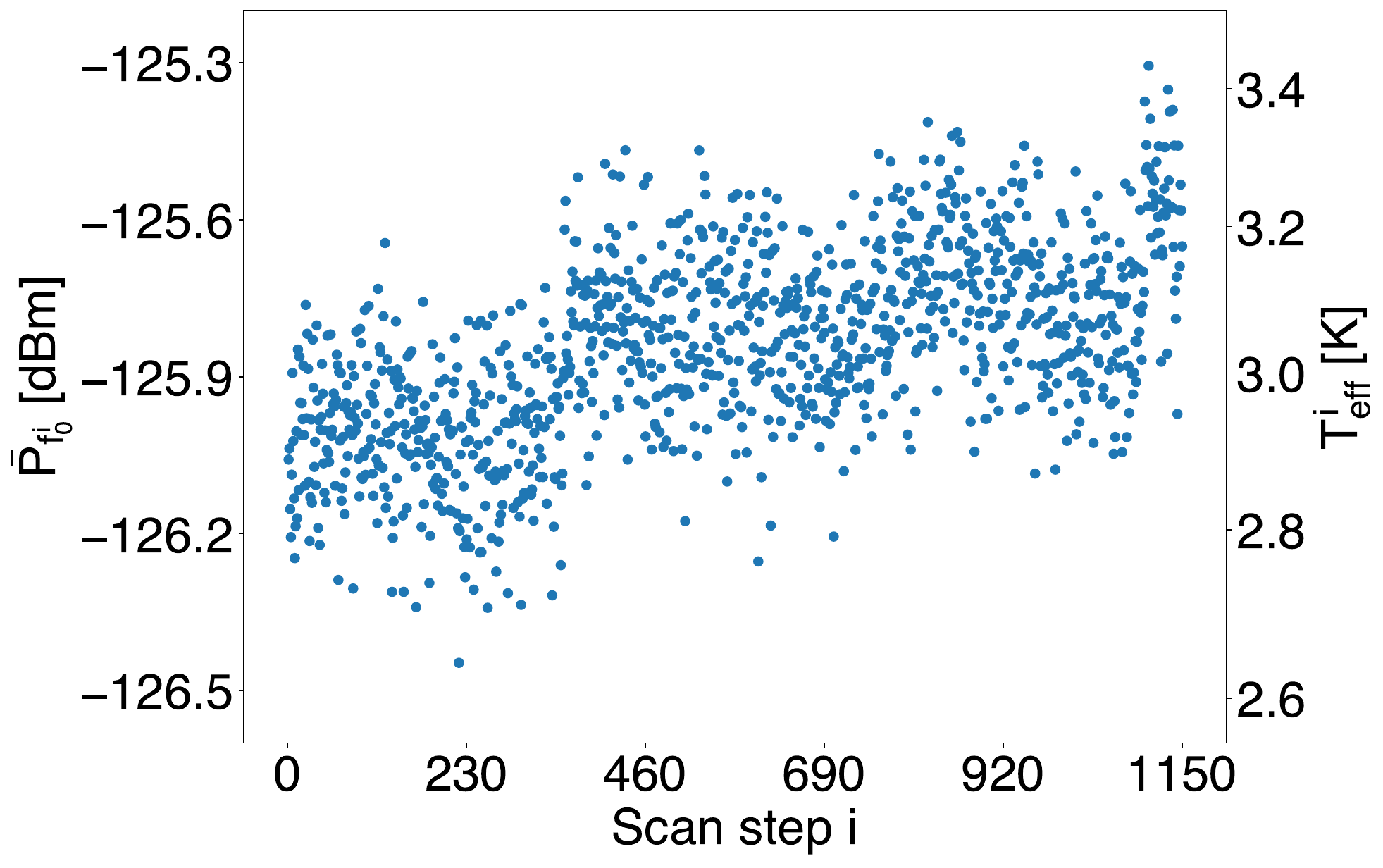} 
    \caption{Left: The values of $\bar{P}_{f_0^i}/\sigma_{f_0^i}$ for each scan. Right: The power $\bar{P}_{f_0^i}$ and the corresponding effective temperature $T^i_{\rm eff}$ for each scan.
     }
    \label{rawdD}
\end{figure}
The scanning experiment unfolded over two sessions: the first covering 350 resonant frequencies from April 12-14, 2023, and the second completing 800 measurements from April 23-26, 2023.
We present the recorded data obtained during the DPDM searches, with a total of $N_{\rm bin} = 1150$ scans (labeled $i=1,2,...,N_{\rm bin}$). For each scan, only the bin at the resonant frequency $f_0^i$ is taken into account in the data analysis. Each bin consists of $N=t_{\rm int}\Delta f_0 = 1150$ samples of measured power, corresponding to an integration time of $t_{\rm int}=100\,{\rm s}$ and a bin size of $\Delta f_0=11.5\,{\rm Hz}$. We defined the sample average as $\bar{P}_{f_0^i}$ and its standard deviation as $\sigma_{f_0^i}$. The recorded power is a two-point correlation function of voltage. The Gaussian nature of the voltage fluctuation leads to a chi-squared distribution with $2$ degrees of freedom of the measured power, satisfying $\bar{P}_{f_0^i}\approx \sigma_{f_0^i}$. We show values of $\bar{P}_{f_0^i}/\sigma_{f_0^i}$ as a function of scan step $i$ in the left panel of Fig.\,\ref{rawdD}. Their distribution is centered on $1$ and follows a Gaussian distribution due to the central limit theorem.

We convert the power into the corresponding effective temperature $T_{\rm eff}^i$ of each scan through $T^i_{\rm eff} \equiv {\bar{P}_{f_0^i}}/{(G_{\rm net} k_b \, \Delta f_{0})}$. The right panel of Fig.\,\ref{rawdD} shows their distribution, ranging from $2.6\,{\rm K}$ to $3.4\,{\rm K}$, with a mean of $3.0\,{\rm K}$. The variation is suppressed by a factor of $\sqrt{N}\approx 34$ compared to typical values of $\sigma_{f_0^i}$. 
A slight upward trend in the distribution is observed, occurring at the juncture between two scan sessions. This minor deviation may result from environmental factors, such as helium pressure fluctuations, which generally maintain stability over several days.
We addressed this effect by grouping every $50$ adjacent bins and performing a constant fit, and estimated its deviation:
\begin{equation}
\bar{P} \equiv \frac{\sum_i \bar{P}_{f_0^i}/{\sigma_{f_0^i}^2}}{{\sum_i 1/{\sigma_{f_0^i}^2}}},\qquad \sigma_{\bar{P}}^2 \equiv \frac{1}{49} \sum_i \left( \bar{P}_{f_0^i} -\frac{1}{50}\sum_j \bar{P}_{f_0^j} \right)^2,
\end{equation}
where $\sigma_{\bar{P}}$ is the sample standard deviation of $\bar{P}_{f_0^i}$ subtracted from $\bar{P}$. Contributions to $\bar{P}$ encompass thermal noise in the cavity, amplifier noise, and injection of room temperature via the $30$\,dB attenuator. We then define the normalized power excess as
\begin{equation}
\delta_i \equiv (\bar{P}_{f_0^i}-\bar{P})/\sigma_{\bar{P}}.
\end{equation}
The histogram of normalized power excess in Fig.\,2 of the maintext is well-modeled by a Gaussian distribution with no observed deviations greater than $3\,\sigma$.

The application of constant fit and normalized excess introduces an attenuation factor $\eta_{\rm fit}$ to the potential signal. In our case, there are $50$ resonant bins in a group, and so any signal entering the fit function $\bar{P}$ will be reduced by a factor of $1/50$, resulting in $\eta_{\rm fit}=98\%$. Moreover, the sample standard deviation of $\bar{P}_{f_0^i}$ minus $\bar{P}$, denoted by $\sigma_{\bar{P}}$, can be increased by a fractional uncertainty from the signal. However, we can neglect this effect when conservatively constraining.

In addition to the uncertainty $\sigma_{\bar{P}}$ that characterizes the fit function $\bar{P}$, uncertainties in calibrated parameters could also contribute to biased estimates of DPDM-induced signals. To address these variances, we introduce dimensionless signals $p_i$:
\begin{equation}
p_i\equiv  \frac{\bar{P}_{f_0^i}-\bar{P}}{P_{\rm ref}^i},\qquad P_{\rm ref}^i \equiv \eta_{\rm bin} \eta_{\rm fit}  G_{\rm net} P_{\text{sig}}(m_{A'},f_0^i,\epsilon=1),\end{equation}
where the reference signal power $P_{\rm ref}^i$ takes into account both the microphonics-induced efficiency $\eta_{\rm bin}$ and the attenuation factor $\eta_{\rm fit}$. $P_{\text{sig}}$ is the signal power,
\begin{equation}
    P_{\rm sig}=\frac{1}{4}\, \epsilon^2 \, \frac{\beta}{\beta+1} \,  V\, \frac{C}{3}\, m_{A^\prime}^2\, \rho_{A^\prime}\,\mathcal{F}(m_{A^\prime}, f_0),
    \label{p_sig}
\end{equation}
where $\mathcal{F}(m_{A^\prime}, f)$ is the normalized frequency spectrum of DPDM, see the following section for detail. Now we can use the error propagation formula:
\be  \sigma_{\mathcal{N}}^2= \sum_j \left(\frac{\partial \mathcal{N}}{\partial x_j}\right)^2\sigma_{x_j}^2,\ee
where $\mathcal{N}(x_1,x_2...)$ is a physical quantity that is a function of measured observables $x_1, x_2 \cdots x_j$, to calculate the variance of the dimensionless signals $p_i\equiv (\bar{P}_{f_0^i}-\bar{P})/P_{\rm ref}^i$. This leads to
\be \begin{aligned}
    \sigma_{p_i}^2 = &\left(\frac{\partial p_i}{\partial \bar{P}_{f_0^i}} \sigma_{\bar{P}}\right)^2 + \sum_j \left(\frac{\partial p_i}{\partial P_{\rm ref}^i} \frac{\partial P_{\rm ref}^i}{\partial j} \sigma_j  \right)^2 \\
    = & \left( \frac{\sigma_{\bar{P}}}{P_{\rm ref}^i} \right)^2 \left(1 + \delta_i^2 \sum_j F_j^2 \right).
    \end{aligned}
    \ee
where the sum of calibrated parameters includes $j \in { V_{\rm eff}, \beta, G_{\rm net} }$, and the fractional uncertainties $F_j$ are defined as
\be \sum_j F_j^2 \equiv \sum_j \left( \frac{\partial P_{\rm ref}^i}{\partial j} \frac{\sigma_j}{P_{\rm ref}^{i}} \right)^2 =  \frac{\sigma_{V_{\rm eff}}^2}{V_{\rm eff}^2}+\left(\frac{1}{\beta}-\frac{1}{1+\beta}\right)^2\sigma_{\beta}^2+\frac{\sigma_{G_{\text{net}}}^2}{G_{\text{net}}^2}.\ee
The fractional uncertainties $F_j$ corresponding to the numerical uncertainties $\sigma_j$ are presented in Table\,I of the maintext. We note that the dimensionless amplification factor $G_{\text{net}}$ is taken as $(56.52 \pm 0.08)$\,dB$\rightarrow 10^{(5.652 \pm 0.008)}$ in this calculation.

Since the reference signal $P_{\rm ref}^i$ is defined in terms of $\epsilon = 1$, we can directly express the probability function in terms of $\epsilon$. Specifically, we have
\begin{equation}
    {\rm Pr} \left(p_i| \epsilon, m_{A'} \right) = \prod_i \frac{1}{\sqrt{2\pi}\sigma_{p_i}} \exp \left( -\frac{(p_i-\epsilon^2)^2}{2\sigma_{p_i}^2} \right)/ {\rm Const},
   \label{likelihood}
\end{equation}
where ${\rm Const}$ is a normalization factor that is irrelevant to the calculation. In principle, the probability function is the product of different resonant bins $i$. However, in practice, due to the fact that each scan was separated by $\approx m_{A^\prime}/(2\pi Q_{\rm DM})\approx 1.3$\,kHz and the narrow bandwidth $\Delta f_0$ compared to it, the products in Eq.\,(\ref{likelihood}) only consider the two nearby resonant bins.

To obtain the $90\%$ upper limit on the kinetic mixing coefficient $\epsilon$ for a given DPDM mass $m_{A^\prime}$, we inversely solve the equation
\begin{equation}    \frac{\int_0^{\epsilon_{90\%}^2} {\rm Pr} \left(p_i|\epsilon, m_{A'} \right)\, {\rm d}\epsilon^2}    {\int_0^{\infty} {\rm Pr} \left( p_i| \epsilon, m_{A'} \right)\, {\rm d}\epsilon^2} = 90\%,
\end{equation}
for $\epsilon_{90\%}$, and the results are displayed in Fig.\,3 of the maintext.

\section{Dark Photon Dark Matter and Cavity Response}
In this section, we review the profile of dark photon dark matter (DPDM) used in the maintext, and presents how a cavity mode is excited by DPDM or thermal noise. 
The standard halo model assumes that dark matter has undergone the process of virialization, which balances the gravitational potential energy of the system with the kinetic energy of its components. In the laboratory frame, the local dark matter velocity distribution satisfies
\begin{equation}
    \mathcal{F} (\vec{v}\,)= \left(2\pi \vvir^2/3\right)^{-3/2} \, \exp \left(- \frac{3(\vec{v}-\vec{v}_g)^2}{2 \vvir^2} \right),
    \label{v_dis}
\end{equation}
where $\vvir \approx 9 \times 10^{-4} c$ is the virial velocity in terms of the speed of light, and $|\vec{v}_g| \approx (2/3)^{1/2} \vvir$ is the Earth velocity in the galactic frame \cite{Turner:1990qx,Lacroix:2020lhn}.

Bosonic dark matter with a mass below $\mathcal{O}(1)$ eV exhibits wave-like properties due to their high occupation number. The frequency of non-relativistic wave-like dark matter is $2\pi f\approx m_{\rm DM}(1+v^2/2)$. The distribution of this frequency comes from Eq.\,(\ref{v_dis}) and can be well approximated by \cite{Brubaker:2017rna, Cervantes:2022epl} 
\begin{equation}
    \mathcal{F}_{\rm DM}(f)= 2 \left( \frac{f-f_{\rm DM}}{\pi} \right)^{1/2}
   \left( \frac{3}{1.7 f_{\rm DM} \vvir^2 } \right)^{3/2} \exp \left( -\frac{3(f-f_{\rm DM})}{1.7 f_{\rm DM} \vvir^2} \right),\label{Ff}
\end{equation}
where $2 \pi f_{\rm DM} \equiv m_{\rm DM}$ corresponds to the frequency of rest mass $m_{\rm DM}$. Eq.\,(\ref{Ff}) satisfies the normalization condition 
$\int \mathcal{F}_{\rm DM}(f)\, \D f =1$. 

For DPDM searches in this study, the signal power spectral density (PSD) in Eq.\,(1) of the maintext depends on the two-point correlation function of DPDM wavefunctions. Under the Lorenz gauge condition $\partial_\mu A^{\prime \mu}=0$, we can determine the DPDM correlation from both the local energy density of dark matter, i.e., $\langle \partial_t \vec{A}^{\prime} \cdot \partial_t \vec{A}^{\prime *} \rangle =\rho_{\rm DM}$ and the energy distribution in Eq.\,(\ref{Ff}), yielding:
\begin{equation}
    \langle \vec{A}^{\prime}(f) \cdot \vec{A}^{\prime*}(f^{\prime})\rangle=
    \rho_{\rm DM}\, \mathcal{F}_{\rm DM}\left(f \right)  \delta\left(f-f^{\prime}\right) / \left( 4\pi^2 \bar{f} f \right),\label{2pDPDM}
\end{equation}
where $\bar{f} \equiv \int f \mathcal{F}_{\rm DM}(f) \D f  \approx m_{A^{\prime}}/(2\pi)$ is the average frequency. Here $\langle \cdots \rangle$ denotes the ensemble average of DPDM fields. These fields can be described as an incoherent superposition of individual vector fields \cite{Guo:2019ker,Chen:2021bdr}. The directions of the DPDM fields are thus isotropically distributed, resulting in a factor of $1/3$ for the correlation between DPDM fields along a given axis, compared to Eq.\,(\ref{2pDPDM}).

We next consider the Maxwell equation coupled with the effective current induced by DPDM, i.e., $\vec{J}_{\text{eff}} = \epsilon\, m_{A^\prime}^2 \vec{A}^{\prime}$,
\begin{equation}
    \Box \vec{E}(t,\vec{x})=\epsilon\, m_{A^{\prime}}^2\,\partial_t\vec{A'}(t,\vec{x}).
    \label{helm}
\end{equation}
The boundary condition of a cavity decomposes the electric field into a discrete sum of orthogonal cavity modes
\be \vec{E} (t,\vec{x})=\sum_n   e_n(t) \vec{E}_n (\vec{x}),\label{Edecom}\ee
where $e_n(t)$ parameterize time-evolution of each mode. $\vec{E}_n (\vec{x})$ form a complete and orthogonal basis within the cavity
\be \nabla^2\vec{E}_n +(2 \pi f_n)^2\vec{E}_n=0, \qquad \int \D V \vec{E}_n \cdot \vec{E}_m^*=\delta_{mn} \ee
with resonant frequency labelled by $f_n$.
Note that our calculation is conducted in the interaction basis, treating the dark photon as an effective current that sources the electromagnetic field. Alternatively, calculations can be performed in the mass basis, employing the boundary condition associated with the screening effect, where both the dark photon and photon fields are coupled to the electromagnetic current~\cite{Chaudhuri:2014dla}.

One can take Eq.\,(\ref{Edecom}) into Eq.\,(\ref{helm}), and project it with the ground mode $\vec{E}_0$ that has the largest overlapping with DPDM. The equation of motion in the frequency domain becomes
 \begin{equation}
    \left(f^2- f_0^2- \mi \frac{f  f_0}{Q_{L}}\right)  e_0 (f)=\frac{\mi \, \epsilon \, m_{A^{\prime}}^2 \,  f}{2\pi}  \int \vec{E}_0 \cdot \vec{A^{\prime}} (f)\, \D V+ \sqrt{\frac{2 f_0}{Q_0}} \, f \, {u}_0(f),
     \label{eom_freq}
\end{equation}
where we take into account the cavity energy loss and dissipation due to intrinsic loss, characterized by the load quality factor $Q_L$ and intrinsic quality factor $Q_0$, respectively.
The last term in Eq.\,(\ref{eom_freq}) is the contribution of thermal noise $u_0 (f)$, which arises due to the fluctuation-dissipation theorem.
The two-point correlation function of thermal noise is
\begin{equation}
    \langle u_0(f) u_0^*(f^{\prime}) \rangle= f \,n_{\rm occ}\, \delta(f- f^{\prime}),
\end{equation}
where $n_{\rm occ} \approx k_b T/(h f)$ is the thermal occupation number, and $h$ is the Planck constant.

The energy stored in the cavity mode, i.e., $U_0 \equiv \iint \D f \D f^\prime \langle e_0(f) e_0 (f^\prime)\rangle$ come directly from Eq.\,(\ref{eom_freq}), which contains both the signal and thermal noise $U_0=U_{\rm sig}+U_{\rm th}$. The signal part is
\begin{equation}
\begin{aligned}
    U_{\rm sig}
    &=\epsilon^2 m^3_{A^{\prime}} \iint \frac{\D f \D f^{\prime}}{8 \pi^3} \frac{f \,\rho_{\rm DM}\,\mathcal{F}_{\rm DM}\left(f \right)\, \delta(f-f^{\prime})}{(f^2-f_0^2)^2+\left( f f_0/Q_L\right)^2}\, V \,\frac{C}{3} 
    \\
    &\approx\frac{Q_L}{8\pi f_0} \epsilon^2 m_{A^{\prime}}^2\, V\, \frac{C}{3}\,  \rho_{\rm DM}\,\mathcal{F}_{\rm DM}(f_0),
\end{aligned}
\end{equation}
where we assume that the cavity response width $f_0/Q_L$ is much narrower than the DPDM width of   $f_{A^{\prime}}/10^6$ to simplify the expression.
The DPDM correlation in Eq.\,(\ref{2pDPDM}) is used, rendering the form factor
\begin{equation}
\begin{aligned}
    C &\equiv \frac{3}{V} \langle \left\vert {\int}_{V} \, \vec{E}_0 \cdot \hat{A^\prime}\, \D V\right\vert^2 \rangle \\
    &=\frac{3}{V} \int \frac{\D^2 \Omega}{4\pi}  \left\vert \int \D V   \vec{E}_0 \cdot \hat{\Omega} \right\vert^2 \\
    &= \frac{1}{V} \left\vert {\int}_{V} \, \vec{E}_0 \, \D V\right\vert^2
\end{aligned}
\end{equation}
for randomized DPDM.

The signal power is read from an antenna from the cavity, which takes the form
\begin{equation}
    P_{\text{sig}} =  \frac{\beta}{\beta+1}\, \frac{2 \pi f_0 }{Q_L}\, U_{\rm sig} =\frac{1}{4}\epsilon^2  \, \frac{\beta}{\beta+1}\,  V\, \frac{C}{3} \, m_{A^\prime}^2\, \rho_{\rm DM}\,\mathcal{F}_{\rm DM}(f_0),
\end{equation}
where $\beta/(\beta+1)$ is the fraction of energy delivered into the antenna in terms of the total energy loss of the cavity.

Similarly, the power of thermal noise is
\begin{equation}
    P_{\rm th}= \frac{\beta}{\beta+1} \frac{2\pi f_0}{Q_L} \frac{2f_0}{Q_0} \int \D f \D f^{\prime} \frac{f f^{\prime}\,\langle u_0(f) u_0^*(f^{\prime}) \rangle}{(f^2-f_0^2)^2+\left( f f_0/Q_L\right)^2} =2\pi\,\frac{\beta}{\beta+1}\, k_b T  \frac{f_0}{Q_0}.
\end{equation}

The noise from the amplifier is proportional to its effective noise temperature $T_{\rm amp}$,
\begin{equation}
    P_{\rm amp}= k_b T_{\rm amp}\, \Delta f_{0}.
    \label{Pamp}
\end{equation}
The spectrum of the amplifier noise is flat within a frequency bin $\Delta f_0$, which is taken to be the resonant frequency stability range in this study and is larger than $f_0/Q_0$. Consequently, the amplifier noise dominates over the thermal noise when $T_{\rm amp}\approx T$.

\end{document}